\documentclass[nofootinbib]{revtex4}
\usepackage{graphicx}
\usepackage{latexsym}
\def\be{\begin{equation}}
\def\ee{\end{equation}}
\def\bea{\begin{eqnarray}}
\def\eea{\end{eqnarray}}

\begin{document}

\hfill USTC-ICTS-17-09

\title{Direct couplings of mimetic dark matter and their cosmological effects}

\author{Liuyuan Shen$^{1,2}$}
\author{Yicen Mou$^{3}$}
\author{Yunlong Zheng$^{1,2}$}
\author{Mingzhe Li$^{3}$}
\email{limz@ustc.edu.cn}
\affiliation{$^{1}$Department of Physics, Nanjing University, Nanjing 210093, China}
\affiliation{$^{2}$Joint Center for Particle, Nuclear Physics and Cosmology, Nanjing 210093, China}
\affiliation{$^{3}$Interdisciplinary Center for Theoretical Study, University of Science and Technology of China, Hefei, Anhui 230026, China}

\begin{abstract}
The original mimetic model was proposed to take the role of dark matter. In this paper we consider possible direct interactions of the mimetic dark matter with other matter in the universe, especially the standard model particles such as baryons and photons. By imposing shift symmetry, the mimetic dark matter field can only have derivative couplings. We discuss the possibilities of generating baryon number asymmetry and cosmic birefringence in the universe based on the derivative couplings of the mimetic dark matter to baryons and photons. 
\end{abstract}

\maketitle

\hskip 1.6cm PACS number(s): 98.80.Cq \vskip 0.4cm

\section{Introduction}

The proposed mimetic model \cite{Chamseddine:2013kea}  was first considered as an extension of general relativity, in which the physical metric $g_{\mu\nu}$ is constructed in terms of an auxiliary metric $\tilde{g}_{\mu\nu}$ and a scalar field $\phi$, as follows
\be\label{way}
g_{\mu\nu}=\tilde{g}_{\mu\nu}\tilde{g}^{\alpha\beta}\phi_{\alpha}\phi_{\beta}/M_1^4~,
\ee
where $\phi_{\alpha}\equiv \nabla_{\alpha}\phi$ denotes the covariant derivative of the scalar field with respect to spacetime coordinates, $M_1$ represents a certain mass scale so that the metric is dimensionless. Here we also used the convention of most negative signature for the metrics. Usually when discussing the mimetic model in the literature, people used the convention of unit reduced Planck mass $M_p^2=1/(8\pi G_N)=1$ and only considered the case where $M_1=M_p=1$. In this paper we will show these mass scales explicitly and consider the general cases where $M_1$ is a different scale from $M_p$.
In terms of this transform (\ref{way}) the conformal mode of gravity is isolated to the scalar field in a covariant way and the physical metric is invariant under the Weyl rescalings of the auxiliary metric. The resulted equations by varying the Einstein-Hilbert action, which is constructed from the physical metric $g_{\mu\nu}$, with respect to the auxiliary metric and the scalar field contain the Einstein equation and an equation of motion for an extra scalar mode which can mimic the cold dark matter in the universe, so it is dubbed the mimetic dark matter \cite{Chamseddine:2013kea}.
Later, as shown in Refs. \cite{Golovnev:2013jxa,Barvinsky:2013mea,Hammer:2015pcx}, without introducing the auxiliary metric the mimetic matter can also be considered as a new scalar component with the constraint $\phi_{\mu}\phi^{\mu}=M_1^4$. Such a constraint can be realized by a Lagrange multiplier $\lambda$ in the action, that is, the action can be written as,
\be\label{original}
S=\int d^4x\sqrt{g}[\frac{M_p^2}{2}R+\lambda(\phi_{\mu}\phi^{\mu}-M_1^4)]+S_m~,
\ee
where $g=-{\rm det}|g_{\mu\nu}|$ and $S_m$ is the action for other matter in the universe. These two points about the mimetic model are equivalent, at least classically. We will take the second point in this paper, as have done in most papers on the mimetic model.

The mimetic model was generalized in Ref. \cite{Chamseddine:2014vna} to include a potential $V(\phi)$, so that the mimetic matter obtains a pressure $p=-V(\phi)$. This is very similar to the generalization in Ref. \cite{Lim:2010yk} for the dusty fluid model. With a potential, the mimetic model has many applications in cosmology, e.g., it can provide inflation, bounce, dark energy, and so on. This stimulated many interests in the literature, for instances, its relation with disformal transformations \cite{Bekenstein:1992pj} were discussed in Refs. \cite{Deruelle:2014zza,Arroja:2015wpa,Domenech:2015tca}, the Hamiltonian analyses were given in Refs. \cite{Malaeb:2014vua}, it has been applied in various modified gravity models \cite{Chaichian:2014qba,Nojiri:2014zqa,Leon:2014yua,Astashenok:2015haa,Myrzakulov:2015qaa,Odintsov:2015cwa,Rabochaya:2015haa,
Odintsov:2015wwp,Arroja:2015yvd,Cognola:2016gjy,Nojiri:2016ppu,Odintsov:2016oyz}, and has also been studied with extensive cosmological and astrophysical interests \cite{Saadi:2014jfa,Mirzagholi:2014ifa,Matsumoto:2015wja,Momeni:2015aea,Ramazanov:2015pha,Myrzakulov:2015kda,Astashenok:2015qzw,
Chamseddine:2016uyr,Nojiri:2016vhu,Babichev:2016jzg,Chamseddine:2016ktu,Sadeghnezhad:2017hmr}. The instability problem about the cosmological perturbations of the mimetic model with higher derivatives was recently studied in Refs. \cite{Ijjas:2016pad,Firouzjahi:2017txv,Hirano:2017zox,Mingzheli,Cai:2017dyi,Cai:2017dxl,Takahashi:2017pje}. Some other recent progresses on this topic can be found in \cite{Sebastiani:2016ras,Vagnozzi:2017ilo,Gorji:2017cai}.

In this paper, we will consider the direct couplings of mimetic field to other matter in the universe, especially the couplings to baryons and photons in the universe.
We will concentrate on the original mimetic model \cite{Chamseddine:2013kea} where the mimetic field has the action (\ref{original}) and mimics the dark matter in the universe. This model has the shift symmetry, i.e., the action is invariant under the field shift $\phi\rightarrow \phi+C$ by a constant $C$. We think that it is the shift symmetry that excluded the potential $V(\phi)$ and guaranteed the mimetic field to be the dark matter. In addition, the shift symmetry prevented any interactions other than derivative couplings of the mimetic dark matter to the rest of the world. In this paper we will study the possible derivative couplings the mimetic matter can have to the baryons and photons and their corresponding effects in cosmology.

This paper is organized as follows. In Section II, we introduce briefly some properties of the mimetic dark matter with a derivative coupling to the matter current. Then in Section III, we will apply it to the baryogenesis model based on the coupling to baryons. The derivative couplings to photons and their implications in cosmology will be investigated in Section IV. Finally we will conclude in Section V. 

\section{Mimetic dark matter with derivative coupling}

When considering the direct interaction between the mimetic field and other matter content, the original action (\ref{original}) becomes
\be\label{action}
S=\int d^4x\sqrt{g}[\frac{M_p^2}{2}R+\lambda(g^{\mu\nu}\nabla_{\mu}\phi\nabla_{\nu}\phi-M_1^4)]+\int d^4x \sqrt{g} \mathcal{L}_{int}+S_m~.
\ee
As we mentioned in the previous section, the imposed shift symmetry requires the interacting Lagrangian density $\mathcal{L}_{int}$ can only depend on the derivatives of $\phi$.
Its simplest case is given by the following operator, 
\be\label{action1}
\mathcal{L}_{int}=\frac{1}{M_2}\nabla_{\mu}\phi J^{\mu}~,
\ee
here $J^{\mu}$ is the matter current associated with a certain quantum number \footnote{Similar interaction has been proposed in Ref. \cite{Vagnozzi:2017ilo}, there $J^{\mu}=\rho u^{\mu}$ is the energy flux current of the matter and is different from the one considered here.}.
The operator $\nabla_{\mu}\phi J^{\mu}$ has the mass dimension of five, so we need to introduce another mass scale $M_2$ to suppress it. The matter current is inversely proportional to the tensor density $\sqrt{g}$, for example, in a system of particles, the current can be defined as \cite{weinberg}, 
\be
J^{\mu}=\frac{1}{\sqrt{g}}\sum_n q_n\int dx^{\mu}_n \delta^4(x-x_n)~,
\ee
where $q_n$ is the charge taken by the n$th$ particle, $x^{\mu}_n$ is its coordinate, and $\delta^4(x-x_n)$ is the Dirac delta function. So the coupling term in the action is independent of the metric and has no contribution to the total energy-momentum tensor, which is the same as that of the original mimetic dark matter model \cite{Chamseddine:2013kea},
\be
\tilde{T}^{\mu\nu}=2\lambda \nabla^{\mu}\phi\nabla^{\nu}\phi+T^{\mu\nu}~,
\ee
where
\be
T^{\mu\nu}=-\frac{2}{\sqrt{g}}\frac{\delta S_m}{\delta g_{\mu\nu}}~,
\ee
is the energy-momentum tensor of matter.

The gravitational field equation becomes
\be
G_{\mu\nu}\equiv R_{\mu\nu}-\frac{1}{2}Rg_{\mu\nu}=-\frac{1}{M_p^2} \tilde{T}_{\mu\nu}~,
\ee
its trace gives the Lagrange multiplier
\be
2\lambda =-\frac{M_p^2G+T}{M_1^4}~,
\ee
where $G$ is the trace of the Einstein tensor and should not be confused with the Newton constant, $T$ is the trace of the energy-momentum tensor of matter. With this one can see that the energy-momentum tensor of the mimetic field has the form of perfect fluid
\be\label{dm}
T^{\mu\nu}_{\phi}=-(M_p^2G+T)u^{\mu}u^{\nu}~,
\ee
where the four velocity $u^{\mu}=\nabla^{\mu}\phi/M_1^2$, which is normalized as $g_{\mu\nu}u^{\mu}u^{\nu}=1$, this is given by the mimetic constraint, $g^{\mu\nu}\nabla_{\mu}\phi\nabla_{\nu}\phi=M_1^4$.
The energy-momentum tensor (\ref{dm}) showed that the mimetic matter is indeed dust-like, its pressure vanishes and its energy density is given by
\be
\rho_{\phi}=-(M_p^2G+T)~.
\ee
Hence, we see that the mimetic field can mimic the dark matter in the universe and the derivative coupling introduced in the action (\ref{action}) does not change this consequence.

However, this derivative coupling will modify the equation of motion of the mimetic field
\be
2\lambda\Box\phi+2\nabla_{\mu}\lambda \nabla^{\mu}\phi+\frac{1}{M_2}\nabla_{\mu}J^{\mu}=0~,
\ee
where $\Box\equiv \nabla_{\mu}\nabla^{\mu}$. 
This shows that if the current is conserved, such a coupling vanishes and has no effect. However, when coupled to a non-conserved current, one can obtain easily that the above equation is equivalent to
\be
\nabla_{\mu}T^{\mu\nu}=\frac{1}{M_2}\nabla_{\mu}J^{\mu}\nabla^{\nu}\phi~,
\ee
this shows that there will be exchange of energy and momenta between the mimetic dark matter and other matter, this effect is suppressed by $M_2$.

\section{mimetic dark matter and baryogenesis}

In this section we consider the case where the current in Eq. (\ref{action1}) is the baryon current, $J^{\mu}=J^{\mu}_B$. In the standard model of particle physics, the baryon number is conserved at low energy scales or low temperature, but it is violated at high energy scales which were achieved in the early universe. Considering the Friedmann-Robertson-Walker (FRW) universe in which the mimetic field is homogeneous and has non-zero time derivative, $\dot\phi=M_1^2$, the coupling $\nabla_{\mu}\phi J^{\mu}_B/M_2$ reduces to \be
\frac{1}{M_2}\nabla_{\mu}\phi J^{\mu}_B\rightarrow \frac{\dot\phi}{M_2} J^0_B=\frac{M_1^2}{M_2}n_B=\frac{M_1^2}{M_2} (n_b-n_{\bar{b}})~,
\ee
here $n_B$ is the net baryon number density. In such a background with non-vanishing $\dot\phi$, the Lorentz symmetry (more exactly the boost symmetry) is broken and correspondingly the CPT symmetry in the baryon section is violated. Such a CPT violation will make a difference between the baryons and anti-baryons. At the early universe when the baryon number violating processes were in thermal equilibrium, the derivative coupling induces an effective chemical potential for baryons and an opposite one for anti-baryons \cite{Cohen:1987vi}, see also \cite{Li:2001st,Li:2002wd},
\be
\mu_b=\frac{\dot \phi}{M_2}=\frac{M_1^2}{M_2}=-\mu_{\bar{b}}~,
\ee
so they have different thermal distributions and one will get a temperature dependent baryon number density \cite{Kolb:1990vq},
\be
n_B=n_b-n_{\bar{b}}=\frac{g_b\mu_bT^2}{6}~,
\ee
where $g_b=2$ is the number of degree of freedom of the baryon and $T$ is the temperature.
On the other hand, the entropy density of the universe is given by \cite{Kolb:1990vq}
\be
s=\frac{2\pi^2}{45} g_{\ast s} T^3~,
\ee
where $g_{\ast s}$ counts the effective degrees of freedom of the species which contribute to the entropy of the universe, the main contributions come from the relativistic particles. So usually $s$ is approximately at the same order of the number density of radiation.
With this we can obtain the baryon-to-entropy ratio
\be\label{result1}
\frac{n_B}{s}=\frac{15 g_b}{4\pi^2}\frac{\mu_b}{g_{\ast s} T}=\frac{15}{2\pi^2}\frac{M_1^2}{g_{\ast s} M_2T}\sim 10^{-2} \frac{M_1^2}{M_2T}~,
\ee
where we have considered $g_{\ast s}\sim 100$ at the radiation dominated epoch.

This provides a model of producing the baryon number asymmetry thermally. Such kind of models are different from the conventional baryogenesis models where three conditions must be satisfied as first proposed by Sakharov \cite{Sakharov:1967dj}, one of which is the departure from thermal equilibrium. The key point leading to the baryon number asymmetry in the model considered here is the CPT violation. Similar models were proposed in Ref. \cite{Cohen:1987vi} where the scalar field was not determined, in Refs. \cite{Li:2001st,Li:2002wd} where the scalar field was identified with the dark energy, and in Refs. \cite{Davoudiasl:2004gf,Li:2004hh} where the scalar field was related to the curvature scalar. The baryogenesis model in Refs. \cite{Li:2001st,Li:2002wd} gives a picture that the current accelerating expansion of the universe (driven by the dark energy) and the generation of baryon number asymmetry can be described uniformly in a same framework. The model considered here gives another picture that it is the dark matter (the mimetic field) playing an important role in producing the baryon number asymmetry.

The equation (\ref{result1}) shows that the baryon number asymmetry was less at earlier time (higher temperature) and became larger at later time with the decreasing of the temperature. This asymmetry froze out at the temperature $T_D$ when the baryon number violating interactions decoupled from the thermal bath,
\be\label{result2}
\left(\frac{n_B}{s}\right)_D\sim 10^{-2} \frac{M_1^2}{M_2T_D}~,
\ee
after that the baryon number is conserved and the coupling (\ref{action1}) will have no effect on baryons. The baryon number asymmetry in our current universe is about $(n_B/s)_D\sim 10^{-10}$, as required by the big bang nucleosynthesis and the observational data of the cosmic microwave background radiation (CMB) \cite{Ade:2015xua}. On the other hand the decoupling temperature $T_D$ of the baryon number non-conservation is around $100$ GeV known from the standard model, so we have the relation
\be
M_1\sim 10^{-3}\sqrt{M_2 ~{\rm GeV}}~.
\ee
This shows that the scale $M_1$, which quantifies the time derivative of the mimetic field, cannot be quite large. Even the scale $M_2$ is as large as the Planck mass $M_p\sim 10^{18}$ GeV, $M_1$ is around $10^6$ GeV.

The consequence (\ref{result2}) also implies that the baryon isocurvature perturbation produced in this model is negligibly small. Baryon isocurvature perturbation is the spatial fluctuation of the baryon-to-entropy ratio.
The equation (\ref{result2}) shows that this ratio is quite homogeneous, because it is only determined by the scales $M_1$, $M_2$ and the decoupling temperature $T_D$, which is further fixed by the parameters of the standard model.

We can also turn to the leptogenesis model in which we replace the baryon current in (\ref{action1})  with the $B-L$ current, here $L$ denotes the lepton number. In this leptogenesis model, we will have the same result (\ref{result2}) except that $T_D$ is the decoupling temperature of interactions which violate $B-L$ instead of the baryon number. From the viewpoint of particle physics, this decoupling temperature is usually much higher than the electroweak scale, so the scale $M_1$ can be relatively larger.

\section{cosmic birefringence induced by the mimetic dark matter}

In this section we will consider the derivative coupling of the mimetic dark matter to photons via the term (\ref{action1}), in which $J^{\mu}$ constructed from the electromagnetic field is the Chern-Simons current,
\be
J^{\mu}=A_{\nu}\widetilde{F}^{\mu\nu}~,
\ee
where $\widetilde{F}^{\mu\nu}=(1/2)\epsilon^{\mu\nu\rho\sigma}F_{\rho\sigma}$ denotes the dual tensor of the electromagnetic tensor, which is defined as $F_{\mu\nu}=\partial_{\mu}A_{\nu}-\partial_{\nu}A_{\mu}$. The divergence of this current is proportional to the Chern-Pontryagin density,
\be
\nabla_{\mu}J^{\mu}={1\over 2}F_{\alpha\beta}\widetilde{F}^{\alpha\beta}~.
\ee
Such a coupling conserves gauge invariance but induce Lorentz and CPT violations in the photon sector with the non-trivial background of the mimetic field. The effect of the Chern-Simons term is to rotate the polarization directions of photons when they propagating within this backgound \cite{Carroll:1989vb,Carroll:1998zi}. For CMB, the rotation angle is \cite{Li:2008tma}
\be
\chi=\frac{1}{M_2}(\phi_{lss}-\phi_0)~.
\ee
This phenomenon is dubbed cosmic birefringence in the literature. In above equation, the subscript $lss$ means the last scattering surface, at which CMB photons decouples with matter, and $0$ represents the value at present time. Usually this angle is anisotropic and its anisotropy depends on the distribution of $\phi_{lss}$ on the last scattering surface, that means the rotation angle is direction-dependent. However, at the background (at the leading order) the mimetic constraint leads to $\phi=M_1^2 t$, we will have an isotropic rotation angle, 
\be
\chi=\frac{M_1^2}{M_2}(t_{lss}-t_0)~.
\ee
More detailed discussions on the anisotropic rotation can be found in Refs. \cite{Li:2008tma,Kamionkowski:2008fp,Li:2013vga}. In the model considered here, the anisotropy of the rotation angle, which depends on the fluctuation of $\phi_{lss}$, is negligibly small due to the mimetic constraint.
Because $t_{lss}$ is much smaller than $t_0\sim 1/H_0$, here $H_0$ is the current Hubble rate, the rotation angle is about
\be\label{result3}
\chi\sim -\frac{M_1^2}{M_2H_0}~.
\ee
The cosmic birefringence will change the power spectra of CMB polarization \cite{Lue:1998mq,Feng:2004mq,Feng:2006dp} because it will convert part of the E-mode polarization to the B-mode. This result provides us the possibility of measuring the rotation angle in terms of the CMB observational data, as has been done firstly in Ref. \cite{Feng:2006dp}. This has been developed into an important and precise method to test CPT symmetry. Currently the upper limit on the rotation angle is about $\mathcal{O}(1^{\circ})$ \cite{Aghanim:2016fhp} \footnote{in the future, the ground-based CMB experiment AliCPT, located in Ali region of Tibet, China can detect a rotation angle as small as $0.01^{\circ}$ \cite{Li:2017lat,AliCPT}.}.
Using the radian measure, the constraint on the rotation angle is $|\chi|\lesssim 10^{-2}$, with Eq. (\ref{result3}) this requires
\be
M_1\lesssim 0.1\sqrt{M_2 H_0} \sim 10^{-22} \sqrt{M_2~{\rm GeV}}~,
\ee
where we have considered $H_0\sim 10^{-43}$ GeV. This result showed that $M_1$ should be extremely small. For example, if we set $M_2=M_p$, $M_1\lesssim 10^{-4}$ eV. Compared with the requirement of baryogenesis in the previous section, the scale $M_1$ should be much smaller, otherwise the rotation angle or the CPT violating signal would exceed the limit by the observational data.  

One can generalize these discussions to the cases of higher order couplings, i.e., the coupling terms contain higher dimensional operators.  We insist on discussing those interactions conserving the gauge invariance of the electromagnetic field and the shift symmetry of the mimetic dark matter. We also exclude the coupling term breaking the Lorentz covariance explicitly. The next order coupling satisfying these requirements is
the following dimension-$7$ operator,
\bea\label{coupling2}
\mathcal{L}_{int}&=& \frac{\xi}{M^3_2} \phi^{\rho}(\nabla_{\rho} F_{\mu\nu})\widetilde{F}^{\mu\nu}~,
\eea
where $\xi$ is a dimensionless coupling constant which is usually thought to be order one. This coupling term can be translated into the familiar form after some calculations, as showed below,
\bea
\mathcal{L}_{int}&=&\frac{\xi}{M^3_2} \phi^{\rho}(\nabla_{\rho} F_{\mu\nu}){1\over 2}\epsilon^{\mu\nu\alpha\beta}F_{\alpha\beta}\nonumber\\
&=&\frac{\xi}{M^3_2} \phi^{\rho}\nabla_{\rho} (F_{\mu\nu}{1\over 2}\epsilon^{\mu\nu\alpha\beta}F_{\alpha\beta})-\frac{\xi}{M^3_2} \phi^{\rho}F_{\mu\nu}{1\over 2}\epsilon^{\mu\nu\alpha\beta}\nabla_{\rho}F_{\alpha\beta}\nonumber\\
&=&\frac{\xi}{M^3_2} \phi^{\rho}\nabla_{\rho} (F_{\mu\nu}\widetilde{F}^{\mu\nu})-\mathcal{L}_{int}~,
\eea
so that
\be
\mathcal{L}_{int}= \frac{\xi}{2M^3_2} \phi^{\rho}\nabla_{\rho} (F_{\mu\nu}\widetilde{F}^{\mu\nu})\rightarrow -\frac{\xi}{2M^3_2} \Box\phi F_{\mu\nu}\widetilde{F}^{\mu\nu}\rightarrow \frac{\xi}{M^3_2} (\nabla_{\mu}\Box\phi) A_{\nu}\widetilde{F}^{\mu\nu}~,
\ee
where arrows denote equivalences up to some total derivative terms.
This interacting Lagrangian density together with $\sqrt{g}$ is not a topological term and will have a contribution to the energy-momentum tensor which sources the gravitational field. However, it is easy to see that this contribution is proportional to the Chern-Pontryagin density $ F_{\mu\nu}\widetilde{F}^{\mu\nu}$ of the electromagnetic field, this is extremely small during the period from last scattering to now, this period is much later than the radiation dominated epoch, and we can neglect the back-reaction of this term to the spacetime safely.  What we focus on is the rotation angle of CMB polarization induced by this coupling. According to the discussions in the previous section, the rotation angle should be
\be
\chi={\xi\over M^3_2} (\Box\phi_{lss}-\Box\phi_0)=3\xi {M_1^2\over M^3_2} (H_{lss}-H_0)\sim {M_1^2\over M^3_2} H_{lss}~,
\ee
where we have considered the equation
\be
\Box\phi=\ddot \phi+3H\dot\phi=3HM_1^2~,
\ee
and neglected $H_0$ when compared with $H_{lss}\sim 10^{-38}$ GeV.  From the observational constraint on the rotation angle $|\chi|\lesssim 10^{-2}$, we have
\be
M_1\lesssim 10^{18}M_2\sqrt{\frac{M_2}{\rm GeV}}~.
\ee
This is automatically satisfied as long as the scale $M_2$ is above $1$ GeV, because $M_1$ should be less than $M_p$. On the contrary, if $M_2$ is much higher than $1$ GeV, the rotation angle induced by the coupling (\ref{coupling2}) would be too small to be detected. 

Dimension-$8$ operators can be built as $\phi^{\rho}\phi_{\rho}F_{\mu\nu}\widetilde{F}^{\mu\nu}$ and $\phi^{\rho}\phi_{\mu}F_{\rho\nu}\widetilde{F}^{\mu\nu}$. The former one has no effect because the mimetic constraint $\phi^{\rho}\phi_{\rho}=M_1^4$ renders this term to be the Chern-Pontryagin density (except a constant coefficient), which is a total derivative and can be dropped out from the total action. The latter operator is the same as the former one up to a constant factor, as shown in the Appendix. Hence, both the dimension-$8$ operators have null effect.

\section{conclusion}

In this paper, we considered some direct interactions of the mimetic dark matter to baryons and photons in the universe. With the shift symmetry, the mimetic field can only couple to other matter derivatively. These couplings have null or negligible back reactions to the spacetime and will not spoil the successes of the standard cosmological model. 
However the mimetic field has a non-trivial background with constant but non-vanishing time derivative, within this background the Lorentz and CPT symmetries are broken spontaneously. In terms of this feature, we constructed models to generate baryon number asymmetry at thermal equilibrium and the cosmic birefringence in CMB, showed possible links between the dark matter and other observable phenomena in our universe. 

\section{Acknowledgement}

This work is supported in part by NSFC under Grant No. 11422543 and No. 11653002.

\section{Appendix}

Here we will prove that the operator $\phi^{\rho}\phi_{\mu}F_{\rho\nu}\widetilde{F}^{\mu\nu}$ in four dimension spacetime is equal to $\phi^{\rho}\phi_{\rho}F_{\mu\nu}\widetilde{F}^{\mu\nu}$ up to a constant factor. 
For this we expand it as
\be\label{expansion}
\phi^{\rho}\phi_{\mu}F_{\rho\nu}\widetilde{F}^{\mu\nu}=\phi_i\phi^jF_{j0}\widetilde{F}^{i0}+\phi_0\phi^0F_{0k}\widetilde{F}^{0k}+\phi_0\phi^iF_{ik}\widetilde{F}^{0k}+\phi_i\phi^0F_{0k}\widetilde{F}^{ik}+\phi_i\phi^jF_{jk}\widetilde{F}^{ik}~.
\ee
First we can see that the third and forth terms at the right hand side of the above equation vanish. Using the electric and magnetic vector fields $\vec{E}$, $\vec{B}$ and the relations $F_{0k}\sim E_k$, $F_{ij}\sim \epsilon_{ijk}B^k$, $\widetilde{F}^{0k}\sim B^k$, $\widetilde{F}^{ij}\sim \epsilon^{ijk} E_k$, it is easy to show that roughly the third term is
\be
 \phi_0\phi^iF_{ik}\widetilde{F}^{0k}\sim \phi_0 (\nabla\phi\times \vec{B})\cdot \vec{B}=0~, 
 \ee
and the forth term is 
\be
\phi_i\phi^0F_{0k}\widetilde{F}^{ik} \sim \phi^0 (\nabla\phi\times \vec{E})\cdot \vec{E}=0~.
 \ee
The first and fifth terms at the right hand side of Eq. (\ref{expansion}) can be combined as
\bea
& &\phi_i\phi^jF_{j0}\widetilde{F}^{i0}+\phi_i\phi^jF_{jk}\widetilde{F}^{ik}=\phi_i\phi^j\epsilon^{0ikl}(\frac{1}{2}F_{0j}F_{kl}+F_{0l}F_{jk})\nonumber\\
&=& \frac{1}{\sqrt{g}}(\phi^1\phi_1+\phi^2\phi_2+\phi^3\phi_3)(F_{01}F_{23}+F_{02}F_{31}+F_{03}F_{12})\nonumber\\
&=& {1\over 2}\phi^i\phi_i \epsilon^{0kjl}F_{0k}F_{jl}=\phi^i\phi_i F_{0k}\widetilde{F}^{0k}~.
\eea
So, Eq. (\ref{expansion}) is 
\be
\phi^{\rho}\phi_{\mu}F_{\rho\nu}\widetilde{F}^{\mu\nu}=\phi_0\phi^0F_{0k}\widetilde{F}^{0k}+\phi^i\phi_i F_{0k}\widetilde{F}^{0k}=\phi_{\rho}\phi^{\rho}F_{0k}\widetilde{F}^{0k}={1\over 4}\phi_{\rho}\phi^{\rho}F_{\mu\nu}\widetilde{F}^{\mu\nu}~,
\ee
at the last step we have considered 
\bea
F_{\mu\nu}\widetilde{F}^{\mu\nu}=F_{0k}\widetilde{F}^{0k}+F_{k0}\widetilde{F}^{k0}+F_{ij}\widetilde{F}^{ij}
= 2F_{0k}\widetilde{F}^{0k}+F_{ij}\epsilon^{ij0k}F_{0k}=4F_{0k}\widetilde{F}^{0k}~.
\eea

{}

\end{document}